# Freezing-out of heavy isotopes of Kr

A. R. Cholach[a,*], D. V. Yakovin[b], N. I. Latkin[a,b], I. A. Sidorov[a]

[a]Boreskov Institute of Catalysis, Akademik Lavrentiev Ave 5, 630090 Novosibirsk, Russia
[b]Institute of Automation and Electrometry, Akademik Koptyug Ave 1, 630090 Novosibirsk, Russia

**Abstract**

The separation of isotopes of natural Krypton at the gas-liquid and liquid-solid phase interfaces was studied under nonequilibrium conditions using a cryogenic cell and mass spectrometry. Condensate formation during Kr cooling begins at an equilibrium temperature, which corresponds to the partial pressure of the dominant isotope $^{84}$Kr, and is accompanied by depletion of the gas phase $^{84}$Kr with a separation coefficient of ~0.92 due to the excess of the heat of condensation over the heat of dissolution by 70-100 kJ/mol. The formation of a solid phase near the freezing point is accompanied by depletion of the gas phase by heavy isotopes. The separation coefficients $^{86}$Kr, $^{84}$Kr, $^{83}$Kr, $^{82}$Kr and $^{80}$Kr are about 1.06, 1.11, 0.86, 0.86 and 0.80, respectively, after the transition of ~30% of the atoms to the solid phase. Pressure-selective condensation can be used to separate components with close boiling points when distillation and temperature-selective condensation methods are ineffective, and freezing out of heavy isotopes can be used to enrich elements with practically important isotopes.

---

[*]Email: cholach@catalysis.ru

# 1. Introduction

Isotope-enriched compounds are necessary for the introduction of advanced technologies in chemistry, materials science, biology, medicine, nuclear energy and many others [1-9]. Isotope markers are widely used in medicine, biology and in fundamental research. Uranium compounds enriched with the isotope $^{235}$U form the basis of nuclear energy.

Isotope enrichment is associated with the problem of separation of components with similar physicochemical properties. The fractional distillation method is not applicable for the separation of components with close boiling points; therefore, isotope enrichment is carried out using special methods based on the difference in the properties of the target product and impurities by weight, solubility, adsorption and diffusion capacity, etc. The main methods of isotope separation are electromagnetic separation, gas diffusion, gas or liquid thermal diffusion, gas centrifugation, aerodynamic separation, laser separation, chemical enrichment, distillation, electrolysis and photochemical separation. Each method has its advantages and disadvantages, but all methods are characterized by high energy intensity, duration and the need for expensive equipment [10-12]. In this regard, the creation of new, efficient and economical methods of isotope separation is an urgent task of fundamental research in the field of chemistry and chemical technology.

The heat and temperature of the phase transition of a heavy isotope is higher for kinetic reasons than that of a light isotope. For example, the melting point of D$_2$O is $T_m$ = 3.82°C, and the boiling point $T_b$ = 101.42°C [13, 14] This feature underlies the methods of water purification from salts of its fractions enriched with heavy isotopes: "heavy" water turns into ice with slow cooling of liquid water with a natural isotope content, while "light" water remains in the liquid phase [15-17]. The authors proposed a method for deep purification of NF$_3$ from CF$_4$ with boiling points of 144.4 K and 145.2 K, respectively, at the phase interface under conditions that allow condensation of only NF$_3$ and provide the greatest enrichment of the CF$_4$ gas phase [18, 19]. During water purification, ice (final state) is enriched with heavy isotopes [15-17], whereas during separation at the phase interface, gas (initial phase) is enriched with the target component (CF$_4$).

The purpose of this work is to establish the effectiveness of partial condensation and freezing-out methods [15-19] for the separation of isotopes at the boundaries of the gas-liquid and liquid-solid phases on the example of natural Krypton, which is characterized by a wide spectrum and commensurate isotope content. The studies were carried out under cryogenic conditions, taking into account $T_b$ = 119.735 K and $T_m$ = 115.78 K of natural Kr [13].

## 2. Experimental

The work used Kr with a natural isotope content in a 4-liter cylinder with an initial pressure of $P_0$ = 13 MPa at 300 K. A 0.5-liter stainless steel cylinder was used as a cryostat; the cylinder walls were sanded and wrapped in three layers with 0.5 mm thick copper foil with a free "edge" 10 cm below the bottom. The temperature of the cryostat placed in a stainless steel thermostat in the range of 78-300 K was set by cooling the corresponding part of the copper "edge" with liquid nitrogen. The temperature measurement was carried out using a $T$-type thermocouple with an accuracy of 1 K, mechanically attached to the cylinder near the free copper "edge".

Pressure measurement in the range of 0.1 – 20 MPa in the cylinder and the intake system was carried out using ELPG300 and ProControl HE-200/10bar sensors, respectively, with an accuracy of 0.5%. The cylinder valve and the ELPG300 pressure sensor were at room temperature, so the effective pressure $P$ and temperature $T$ differed from the measured values of $P_{exp}$ and $T_{exp}$ as $P_{exp} > P$ и $T_{exp} < T$, respectively. The correction of $P$ and $T$ was carried out in two ways: under the condition $T_{exp} = T$, the value of $P$ was determined by substituting $T_{exp}$ into the literary data $P(T)$ for natural Kr in the temperature range from the beginning of condensation ($T_{cond}$) to $T_b$; under the condition $P_{exp} = P$, the value of $T$ was determined by substituting $P_{exp}$ into the literary data $P(T)$ [20]. The temperature was measured near the condensate, so the condition $T_{exp} = T$ is more reliable. The lower bound of the temperature of the gas-liquid equilibrium region $T > T_b$ was estimated by the excess of the effective pressure over the equilibrium $P > 101.3$ kPa; the upper bound $T < T_{cond}$ was determined by the transition temperature of the dependence $P(T)$ for an ideal gas (1) to the Clausius-Clapeyron dependence (2) with the corresponding Antoine equation [21]:

$$PV = nRT \tag{1}$$

$$\bar{P} = P_0 \exp\left\{\frac{Q}{RT}\right\} \tag{2}$$

$$log_{10}\bar{P} = 8.097018 - 970.806955/T$$

where $P$ and $\bar{P}$ are the actual and equilibrium pressure; $T$ is the temperature; $V$ and $n$ are the volume and number of moles of gas; $Q$ is the heat of condensation; $R$ is the gas constant; $P_0$ is a constant.

The relaxation time (establishment of equilibrium) of the system ($t_r$) is estimated by the diffusion time of Kr atoms at a distance of the characteristic size of the cryostat L = 10 cm [22]:

$$t_r = \frac{L^2}{2D} \tag{3}$$

where $D = \frac{1}{3}v\lambda$ is the self-diffusion coefficient at the average velocity of the molecule $v = \sqrt{\frac{8RT}{\pi m}}$ and the free path length $\lambda = \frac{RT}{\sigma P N_A \sqrt{2}}$, at the Avogadro constant $N_A$ and the gas constant $R$; for an effective cross-section $\sigma = 4\pi r^2$, the atomic radius $r = 1.52$ Å (between the covalent radius 1.16 Å and the Van der Waals radius 2.02 Å [13]) was taken from the experimental value $D = 7.59 \cdot 10^{-6}$ m²/s (273.2 K, 1 atm) [23].

The holding time of the system at each temperature was 5-10 min. The composition of isotopes in gas samples was determined using a monopole mass spectrometer installed in a vacuum chamber. The accuracy of determining the isotopic composition of ~6% was estimated by the standard deviation of the ion current at 15-20 measurements, which did not exceed 3%. The isotope $^{78}$Kr was not taken into account, due to the small content. The separation coefficients, as a result of partial condensation or freezing-out of the natural mixture and enrichment of the gas phase with the $i$-th isotope, were determined by the equation [24]:

$$\alpha_i = \frac{x_i/(1-x_i)}{x_{i,0}/(1-x_{i,0})} \quad (4)$$

where $x_{i,0} = \frac{n_{i,0}}{\sum n_{i,0}}$ and $x_i = \frac{n_i}{\sum n_i}$ are the initial and final contents of the $i$-th isotope with the number of moles $n_{i,0}$ and $n_i$, respectively.

## 3. Results and discussion

The isotopic composition of the initial Kr in Table 1 is close to the literature data on the natural content of isotopes [25].

Table 1. Isotope content (% at) of working Kr (Eq. (4)).

| Isotope | this work | Ref. [25] | $\Delta x_i$ (%) |
|---|---|---|---|
| $^{80}$Kr | 2.247 | 2.29 | -1.8 |
| $^{82}$Kr | 11.490 | 11.62 | -1.1 |
| $^{83}$Kr | 11.253 | 11.53 | -2.4 |
| $^{84}$Kr | 58.330 | 57.20 | 2.0 |
| $^{86}$Kr | 16.680 | 17.36 | -3.9 |

The difference between $P_{exp}$ and $P$ increases by 10-35% (Fig. 1a) and the difference between $T$ and $T_{exp}$ increases by 4.5-6.5 K (Fig. 1b) when the cryostat temperature changes from 220 K to 120 K. Further results used pressure correction (Fig. 1a).

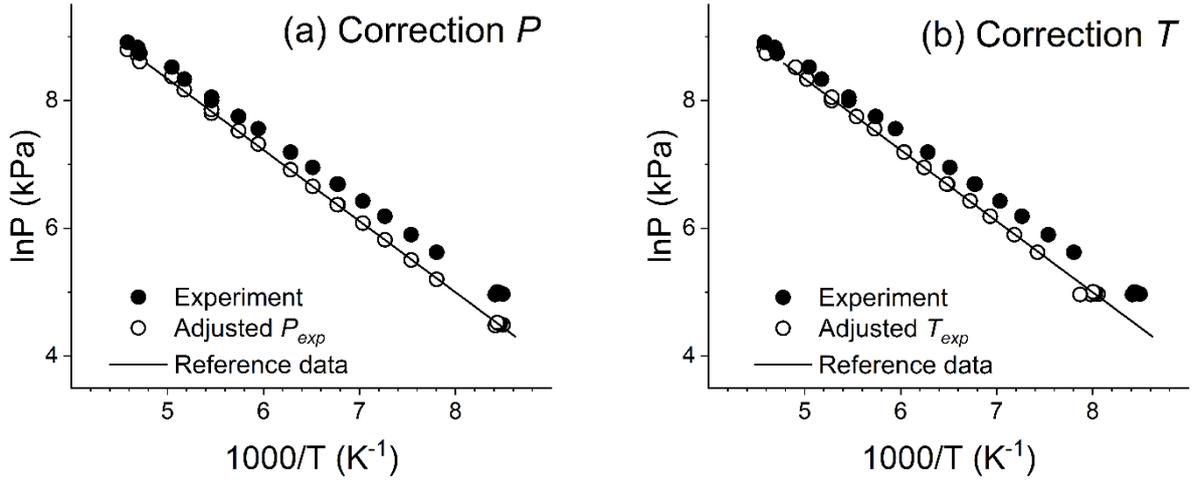

**Fig. 1.** P-T diagram of natural Kr ($P_0$ = 12.6 MPa) in the gas-liquid equilibrium region: initial (filled symbols) and with (a) corrected $P$ at measured $T_{exp}$ and (b) corrected $T$ at measured $P_{exp}$ (blank symbols); the literature data (Eq. (2)) are shown by a line [21].

*3.1. Quasi-equilibrium gas – liquid*

The pressure decreases with a decrease in temperature (Eq. (1)), and the excess of the effective pressure over the equilibrium one $P(T_{cond}) > \bar{P}$ determines the condensation temperature of the gas during its cooling. The values of $T_{cond}$ calculated for total and partial ($P_i$) isotope pressures as effective pressure are given in Table 2, and the dependence of $P(T)$ for natural Kr in the area of calculated $T_{cond}$ is shown in Fig. 2a.

**Table 2.** Total and partial ($P_i$) pressures of Kr isotopes and corresponding values of $T_{cond}$ compared with the experiment.

|  | % at | $P_i$ (kPa) | $T_{cond}$ (K) |
|---|---|---|---|
| Total | 100 | 11075 | 243 |
| $^{80}$Kr | 2.247 | 249 | 133 |
| $^{82}$Kr | 11.490 | 1273 | 165 |
| $^{83}$Kr | 11.253 | 1247 | 165 |
| $^{84}$Kr | 58.330 | 6460 | 217 |
| $^{86}$Kr | 16.680 | 1847 | 175 |
| Fig. 2a |  |  | 217-221 |

The difference in the dependences $P(T)$ during cooling and heating of the cryostat is a consequence of quasi-equilibrium conditions, since the holding time of the system at each temperature of 5-10 min is significantly less than the relaxation time (establishment of equilibrium) $t_r$ = 5-15 h in Fig. 2b, where the gas-liquid features were estimated as follows:

$$n_0 = n_g + n_l; \quad V_0 = V_g + V_l \qquad \text{Material balance}$$

$$n_0 = \frac{P_0 V_0}{RT_0}; \quad n_g = \frac{PV_g}{RT} \qquad \text{Transform Eq. (1)}$$

$$\frac{V_l}{V_0} = \frac{P - dRT}{P - P_0 T/T_0} \qquad (5)$$

$$\frac{n_l}{n_g} = \frac{dRT}{P} \cdot \left(\frac{V_l}{V_0} - 1\right)$$

where $d$ is the temperature-dependent density of the liquid Kr [21]; $n_l$, $n_g$ and $n_0$ are the number of moles of the liquid phase, gas phase and total number of moles; $V_l$, $V_g$ and $V_0$ are the volume of condensate, gas phase and total volume; $T$ and $T_0$ are the actual and initial (room) temperature; $P$ and $P_0$ are the actual and initial (at $T_0$) pressure, respectively.

Table 2 and Fig. 2a show that both during cooling and heating Kr, the linearity termination temperature $P(T)$ indicates partial condensation or boiling of $^{84}$Kr, which prevails in a natural mixture of isotopes (Table 1).

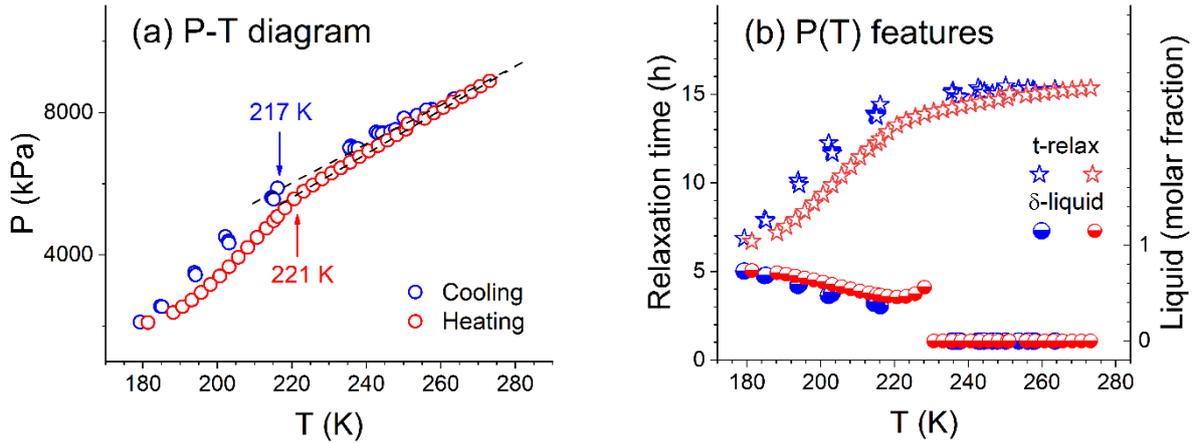

**Fig. 2.** Dependence of (a) $P(T)$ (linearity at higher temperatures is underlined by a dotted line, and the end of linearity is indicated by arrows) and (b) $t_r(T)$ and $\delta_l(T)$ during cooling (blue symbols) and heating (red symbols) of natural Kr in the condensation region (Eq. (3)).

With a relatively low degree of condensation of $n_l/n_g = 0.58$, depletion of the gas phase with the isotope $^{84}$Kr is observed (Table. 3), and the change in atomic fractions $(x_{84,0} - x_{84})/x_{84,0} \sim 0.04$ is significantly higher than the kinetic isotope effect $< 10^{-3}$, taking into account the difference between the system and an ideal gas at high pressures [26]. The high fractionation of isotopes may be due to the excess of the heat of condensation over the heat of dissolution $\Delta Q_{cond}$. Substituting $x_i/x_{i,0}$ into Eq. (2) allows us to estimate the range of values $\Delta Q_{cond} = 65\text{-}96$ J/mol, which is 0.8-1.1% of the heat of condensation of natural Kr $\sim 8.73$ kJ/mol at 212 K [26].

The gas-liquid equilibrium is determined by the equality of the gas flow velocity to the surface and the evaporation rate, but the unlimited solubility of one isotope in the condensate of another leads to forced condensation of any isotope without observing the condition $P_i > \bar{P}$. The flow of isotopes to the surface of the liquid phase is proportional to the natural content of $x_i$ and forms a condensate with the same content, so the rate of evaporation of isotopes from the condensate surface is also proportional to $x_i$. In this regard, the condensation of natural Kr under quasi-equilibrium conditions does not lead to a change in the isotopic composition of practical interest, and the fractionation of isotopes in Table 3 is due only to a slight difference in their thermodynamic properties [26, 27].

**Table 3.** Values of $P_0$; $n_l/n_g$, $V_l/V_0$ and equilibrium $\overline{V_l/V_0}$ in Eq. (5) and $\alpha_i$ in Eq. (4) for natural Kr at different temperatures.

| $T$ | 212.2 K | 204.2 K | 119.3 K | 113.2 K |
| --- | --- | --- | --- | --- |
| $P_0$ (MPa) | 10.34 | 10.63 | 11.99 | 11.99 |
| $n_l/n_g$ | 0.58 | 1.24 | 62.74 | 118.96 |
| $\Delta P/\bar{P}$ | 0.085 | 0.079 | 0.078 | 0.233 |
| $V_l/V_0$ | 0.1123 | 0.1448 | 0.1629 | 0.1610 |
| $\overline{V_l/V_0}$ | 0.0886 | 0.1126 | 0.1626 | 0.1662 |
| $\alpha_{80}/x_{80}$ | 1.04/1.04 | 1.01 | 0.99 | 1.09 |
| $\alpha_{82}/x_{82}$ | 1.06/1.05 | 1.03 | 0.99 | 1.07 |
| $\alpha_{83}/x_{83}$ | 1.05/1.04 | 1.00 | 0.99 | 1.07 |
| $\alpha_{84}/x_{84}$ | 0.92/0.96 | 0.97 | 1.01 | 0.95 |
| $\alpha_{86}/x_{86}$ | 1.06/1.05 | 1.03 | 1.00 | 0.97 |

Pressure-dependent condensation could be expected for isotopes, but not for sure, due to their extremely similar physical properties, with the exception of atomic masses. In fact, isotopes begin to condense as independent components, but the nucleation of the liquid phase returns the isotopic composition to its natural one in accordance with Raoul's law. In addition, the apparent boiling and condensation temperatures of elements with a comparable isotope content (Mg, Cl, Cu, Ga, Rb, etc.) are not a weighted average of partial fractions and temperatures, but refer to the dominant isotope. Partial pressure selective condensation under nonequilibrium conditions can be used to separate components with close boiling points when distillation and temperature-selective condensation methods are ineffective.

*3.2. Quasi-equilibrium liquid – solid*

Unlike the gas-liquid system, the freezing rate of an isotope, without taking into account the influence of crystallization centers, depends only on the temperature of the phase transition, at which the solid phase is thermodynamically more advantageous than the liquid phase, and weakly depends on the content of this isotope.

Holding Kr ($P_0$ = 11.9 MPa) at $T$ = 113±3 K (about $T_m$) for ~1 hour led to depletion of the gas phase by heavy isotopes, which indicates their predominant freezing-out (Table 3). In the condensation region, the ratio of atoms in the liquid and gas phases $n_l/n_g$ increases with decreasing temperature, the actual volume fraction of condensate $V_l/V_0$ is higher than equilibrium $\overline{V_l/V_0}$ (Eq. (5)), and the actual pressure is lower than equilibrium $\Delta P = P - \bar{P}$ due to non-equilibrium conditions. Taking into account the close isotopic composition of the gas and liquid phases and assuming that the observed enrichment or depletion of the gas sample is compensated by the corresponding depletion or enrichment of the solid, the atomic fraction of the $i$-th isotope in the solid phase and the separation coefficients between the gas + liquid and solid phases are estimated in Fig. 3 as follows:

$$x_{i,0} = x_{i,s}\delta_s + x_{i,g}(1 - \delta_s) \qquad \text{Material balance}$$

$$\alpha_{i,s} = \frac{1/x_{i,0} - 1}{\delta_s/(x_{i,0} - x_{i,g}(1 - \delta_s)) - 1} \qquad (6)$$

where $x_{i,0}$, $x_{i,s}$ and $x_{i,g}$ are the atomic fraction of the $i$-th isotope in the initial Kr, solid phase and gas and liquid phases near $T_m$, and $\delta_s = \frac{n_s}{n_0}$ is the molar fraction of the solid phase, where $n_s$ and $n_0$ are the number of moles of the solid and the total number of moles, respectively.

The identity of the measurement method allows us to maintain the degree of deviation of the actual pressure from the equilibrium $\Delta P/\bar{P}$ in the condensation region ($T$ = 119-212 K, Table 3). The difference in density of solid Kr 2.80 g/cm³ [28] and liquid Kr 2.42 g/cm³ [20] at 113 K is not large enough to explain the significant increase in $\Delta P/\bar{P}$ during the formation of the solid phase. The assumption that the excess $\Delta P/\bar{P}$ corresponds to the molar fraction of the liquid phase that has passed into the solid, and the contribution of the solid phase to the actual pressure is ~ $\exp\{-Q_m/RT\}$ = 0.176, where $Q_m$ = 1.64 kJ/mol is the melting heat Kr [29], allows us to estimate $\delta_s$ ~0.32. Then Eq. (6) gives an estimate of the separation coefficients $\alpha_{86,s}$ ~1.06, $\alpha_{84,s}$ ~1.11, $1/\alpha_{83,s}$ ~1.17, $1/\alpha_{82,s}$ ~1.17 и $1/\alpha_{80,s}$ ~1.26, which are comparable to $\alpha$ ~1.26 for the enrichment of ice with deuterium [30] – and also with the difference in the relative velocities of the isotopes Kr (1.2-3.6%) and $H_2O$ and HDO (2.7%), which determine the formation of the solid phase. According to the material balance, the solid phase should not contain $^{80}$Kr, $^{82}$Kr and $^{83}$Kr at $\delta_s$ < 0.09, < 0.06 and < 0.06, with boundary values $\alpha$ ~0.02, ~0.04 and ~0.05, respectively, to compensate for the enrichment of the gas and liquid phases (Eq. (6)).

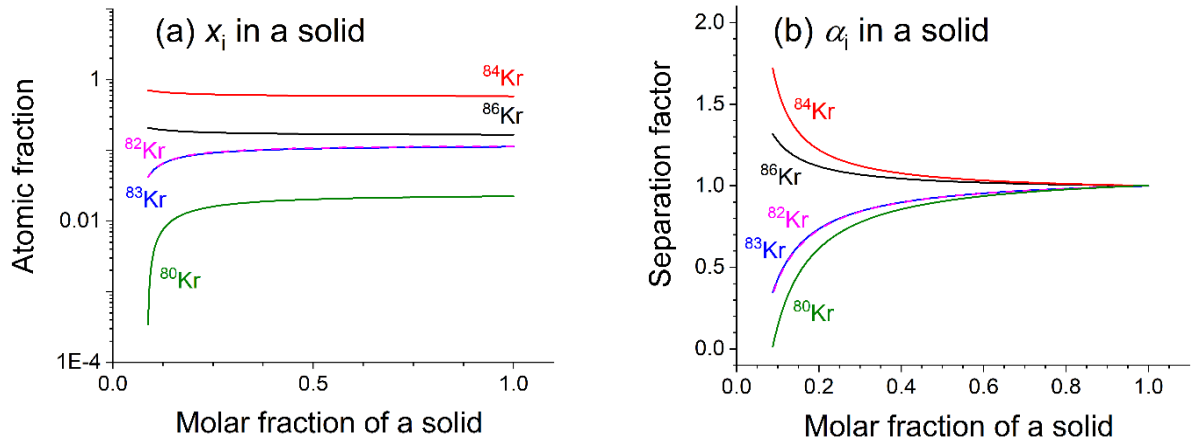

**Fig. 3.** (a) Isotope content ($x_i$) and (b) separation coefficients ($α_i$) for the transition of various fractions of atoms ($δ_s$) to the solid phase (Eq. (6)).

The dependences $x(δ_s)$ and $α(δ_s)$ in Fig. 3 are limited on the left by the complete absence of $^{80}$Kr, $^{82}$Kr and $^{83}$Kr in the solid phase to satisfy the material balance. The predominant freezing-out of heavy isotopes Kr (Fig. 3) and H$_2$O [15-17] can be used to enrich elements with rare practically important isotopes.

## 4. Conclusions

The separation of isotopes of natural Krypton at the gas-liquid and liquid-solid phase interfaces under nonequilibrium conditions has been studied. The formation of the liquid phase upon cooling Kr from the ambient temperature occurs at an equilibrium temperature, which corresponds to the partial pressure of the dominant isotope $^{84}$Kr, and is accompanied by depletion of the gas phase 84Kr with a separation coefficient of ~0.92 due to nonequilibrium conditions and an excess of the heat of condensation over the heat of dissolution by 70-100 kJ/mol. Isotopes begin to condense as independent components, despite their very similar physical properties, but the nucleation of the liquid phase returns the isotopic composition to its natural one in accordance with Raoul's law. Separation can probably be improved by removing the condensate with suitable liquid absorbers or capillaries, thus maintaining non-equilibrium conditions. Pressure-selective condensation can be used to separate components with close boiling points when distillation and temperature-selective condensation methods are ineffective. The retention of Kr of krypton near the freezing point leads to the formation of a solid phase enriched in heavy isotopes. The separation coefficients $^{86}$Kr, $^{84}$Kr, $^{83}$Kr, $^{82}$Kr and $^{80}$Kr are ~1.06, ~1.11, ~0.86, ~0.86 and ~0.80, respectively, after the transition of ~32% of atoms to the solid phase. Preferential freezing-out of heavy isotopes can be used to enrich elements with rare practically important isotopes.

**Acknowledgment**

This work was supported by the Russian Science Foundation (Project #23-23-00033).